# Multiferroicity in the Presence of Exchange Bias: The Case of Spinel $CoMn_2O_4$

P. Kumar [1], P. Das [1], B. K. Kuanr[2], S. Patnaik[1*]

[1]School of Physical Science, Jawaharlal Nehru University, New Delhi 110067, India
[2]Special Centre for Nanoscience, Jawaharlal Nehru University, New Delhi 110067, India
*Corresponding Author: spatnaik@jnu.ac.in

**Abstract:**

Ferrimagnetic spinel materials of formula $AB_2X_4$, where A and B are transition metals and X is oxygen or sulphur, hold promise for the realization of multiferroic characteristics. In this work, we report synthesis of spinel $CoMn_2O_4$ and explore its magnetic, dielectric, and ferroelectric aspects and their correlations. Polycrystalline $CoMn_2O_4$ was synthesized by using the conventional solid-state method. The X-ray diffraction (XRD) and Raman spectroscopy confirmed the phase purity of the synthesized compound. The crystal structure was identified with tetragonal symmetry (I41/*amd* space group). DC magnetization measurements indicate two magnetic transitions: one at temperature $T_1 \sim 186$ K, followed by another Yafet-Kittel (YK) ferrimagnetic transition at $T_2 \sim 86$ K. A frequency independent anomaly in the temperature dependent dielectric permittivity is observed near the low magnetic ordering temperature ($T_2$). This reflects the possibility of the correlation between lattice dynamics and spin ordering in spinel $CoMn_2O_4$. A substantial exchange bias was also observed below $T_2 \sim 86$ K. The change in dielectric permittivity in the presence of applied magnetic field follows the square of the magnetization dependence, which is consistent with Ginzburg-Landau theory. However, the detailed pyroelectric current measurements reveal the absence of intrinsic ferroelectric order.

## 1. Introduction

Multiferroic materials exhibit spontaneous magnetic and polar ordering in a chemically single-phase compound along with coupling between these ordered states. Such materials have become a central focus of contemporary research in condensed matter physics and the development of innovative electronic devices based on such juxtaposed phases of matter [1, 2]. Several potential applications, such as non-volatile memories and ultrasensitive sensors, are envisaged [3, 4]. However, strong magnetoelectric coupling in single-phase systems remains challenging due to the fundamentally conflicting nature of electric and magnetic order parameters [5]. Recent studies show that specific types of magnetic orders, such as helical, spiral, conical, and frustrated spin configurations, can effectively break inversion symmetry and induce ferroelectricity [6, 7]. Notable examples of magnetically induced ferroelectricity include rare-earth manganites such as $RMn_2O_5$ [4], $DyMnO_3$ [8], and $Co_3TeO_6$ [9]. The underlying mechanisms responsible for the emergence of polarization (P) with noncollinear magnetic arrangements can be understood through models based on Dzyaloshinskii-Moriya (DM) interaction [10]. Depending on spin structure, in some cases, the inverse DM interaction is also invoked. This relates to magneto-elasticity induced lattice distortions [11, 12].

The magnetic spinel oxide, $AB_2O_4$ (where A and B are transition metals), are promising materials for studying the magnetoelectric coupling because of their unique and complex magnetic structure. In spinel oxides, A-site cations are enclosed within tetragonal oxygen cages, forming a diamond sublattice, whereas B-site cations occupy the octahedral lattice sites, giving rise to a pyrochlore sublattice, often leading to geometrical frustration [13, 14]. Depending on the nature of the cation, and exchange interactions varied phenomena such as geometrical frustration [15], heavy-fermion behaviour [16], orbital ordering [17], incommensurate magnetic order [18, 19], and quantum phase transitions [20] are reported. Magnetoelectric coupling and multiferroic order has also been studied in chromite spinel $ACr_2O_4$ (A = Co, Mn, Fe, Ni) and [21, 22] and vanadium spinel $AV_2O_4$ (A = Fe and Cd) [23, 24]. However, the relevant study is scarce in manganite spinel $AMn_2O_4$ (A = Zn, Mn, Co, Ni), where manganese ions with multiple valence states can adapt to various crystal field environments and, consequently, facilitate the development of various kinds of exchange interaction. In this series of spinel, $Mn_3O_4$ spinel with the $Mn^{3+}$ ions ($3d^4$) undergoes a Jahn-Teller distortion because of the degeneracy of their $e_g$ orbital, leading to a tetragonal elongation of the crystal lattice (c > a = b) from the cubic (Fd$\bar{3}$m) symmetry [25]. This elongation weakens magnetic exchange interactions along the c-axis and strengthens them within the ab-planes. In general, antiferromagnetic spin chains in ab-planes are formed. However, the orthogonal configuration of spin chains across adjacent planes results in the cancellation of out of plane exchange interactions, so that geometric frustration may not be completely resolved within this lattice symmetry. C. Kemei et al. observed that further orthorhombic distortions can help alleviate this frustration below the Neel temperature $T_N$ [26]. However, the magneto-structural transitions in $Mn_3O_4$ occurs at extremely low temperatures (below 40 K), restricting its practical applications. Moreover, the substitution of various magnetic or nonmagnetic ions at the tetrahedral lattice sites of $Mn_3O_4$ can significantly modify both of its lattice and magnetic structure [27, 28]. In this scenario, $CoMn_2O_4$, has garnered attention due to its ability to enhance the transition temperature while maintaining similar magnetic and structural characteristics. This enhancement is primarily attributed to a decrease in bonding geometry when the $Co^{2+}$ ions occupy the tetrahedral lattice site rather than $Mn^{2+}$ ions, thereby strengthening the Co−O−Mn superexchange interactions [29]. $CoMn_2O_4$ crystallizes in a mixed spinel structure, characterized by distorted tetragonal symmetry. It exhibits two distinct magnetic transitions, typically occurring between 165–190 K and 78–90 K, depending on the synthesis method and size of particles [30, 31]. Some studies report a single magnetic transition near 100 K for $CoMn_2O_4$ single crystal [32]. According to earlier investigations, the lower temperature transition in the range of 78–90 K is attributed to pronounced spin canting, which is a Yafet-Kittel (YK) type spin arrangement similar to $Mn_3O_4$. Here cobalt spin is coupled antiferromagnetically with two manganese spins of pyrochlore lattice, but are slightly canted with respect to each other. The higher temperature transition in the range 165–190 K, on the other hand, is associated with the impurity phase of $Co_{3-x}Mn_xO_4$ [33]. Spinel $CoMn_2O_4$ has been studied for its remarkable electrochemical performance, and for use as electrode materials for the high performance lithium-ion batteries [34, 35]. However, the correlation between its magnetic and lattice degrees of freedom remains unexplored.

Recently, Poojitha et al. highlighted the response of phonon modes near the magnetic transition in $CoMn_2O_4$ through temperature dependent Raman spectroscopy, revealing strong spin-phonon coupling [36]. The spin-phonon coupling in material, where magnetic and lattice order parameters are strongly correlated, can exhibit various properties such as magnetoelectric coupling, the spin-Seebeck effect, the magnetostriction effect, and the thermal Hall effect in multiferroics [37–39]. So, to investigate the interplay between lattice and magnetic order parameters in $CoMn_2O_4$, we synthesized polycrystalline $CoMn_2O_4$ using the conventional solid-state reaction method. The lattice dynamics are studied using dielectric and pyroelectric current measurements across magnetic ordering transitions. Our results provide clear evidence that $CoMn_2O_4$ spinel can exhibit coupling of dielectric and magnetic ordering even in the absence of an external magnetic field. Moreover, the tunability of the dielectric permittivity under an applied magnetic field indicates that $CoMn_2O_4$ offers strong magnetodielectric coupling. These findings lay a compelling foundation for further exploration of $CoMn_2O_4$ as a multifunctional material.

## 2. Experimental Methods

### 2.1. Sample synthesis

Cobalt-manganese spinel, $CoMn_2O_4$, was prepared via the traditional solid-state reaction method by taking analytically pure CoO (99.9%, Sigma-Aldrich) and $Mn_2O_3$ (99.5%, Sigma-Aldrich) as starting materials. These materials were accurately weighed in stoichiometric portions and thoroughly hand-ground in an agate mortar for 5-6 hours to achieve a uniform mixture. The resulting mixture was then calcined in air at 900°C and 1000°C for 12 hours each, with intermittent grindings to improve homogeneity. For further processing, the calcined powder was mixed homogeneously with 5 wt% polyvinyl alcohol (PVA) binder and pressed into circular pellets of 10 mm diameter and 1−1.5 mm thickness through a hydraulic press, followed by sintering at 1100°C for 24 hours.

### 2.2. Experimental characterization

To examine the phase purity and crystal structure, room temperature powder X-ray diffraction (Rigaku Miniflex D/MAX 2500PC, equipped with Cu-$K_\alpha$ radiation, $\lambda = 1.5460$ Å) was performed. These results were further supported by Raman spectra in the wavenumber range of 50 $cm^{-1}$ to 800 $cm^{-1}$, collected by using a Raman spectrometer (Model: Enspectr R 532) with a wavelength of 532 nm. The surface morphology and elemental composition of material was studied with the help of a scanning electron microscope (SEM, JSM-IT200, JEOL) equipped with Energy-Dispersive X-ray spectroscopy (EDX). Temperature and magnetic field-dependent DC magnetic studies of $CoMn_2O_4$ were evaluated with a Vibrating Sample Magnetometer (VSM) installed in a *Cryogenic Physical Properties Measurement System* (PPMS) with temperatures from 1.6 K to 300 K. For the electrical measurements, both surfaces of the sample were polished and then coated with silver epoxy to make good electrical contact. Temperature dependent DC resistivity (measured with two-probe method) and pyroelectric current properties were measured using a 6514 Keithley electrometer. Before pyroelectric current measurement, the sample was poled by cooling it from high to low temperature under an applied electric field. Then at the lowest temperature, the electrodes were short circuited to mitigate the impact of electrostatic stray charges after removing the electric field. P-E loop was measured using a Ferroelectric tester (Precision Premier II - Radiant Technologies, INC.). Dielectric measurements as a function of temperature were carried out using an Agilent E4980A LCR meter with variable frequencies under an AC voltage amplitude of 1 Volt. All temperature dependent electrical measurements were carried out by a controlled programmable temperature controller (Model: 325, LakeShore) associated with a cryogen-free low temperature high magnetic field system.

## 3. Results and Discussion

### 3.1. Structural Analysis

Fig.1(a), depicts the room temperature powder X-ray diffraction (XRD) that was conducted to evaluate the phase purity and crystallographic properties of the synthesized $CoMn_2O_4$. It is reported that $CoMn_2O_4$ crystallizes in tetragonal phase [33]. However, some reports suggest that cubic structure could also be formed with the dominating tetragonal phase [31, 40, 41]. Recently, S. Rajput *et al.* [42], showed splitting of XRD peaks and claimed that a small fraction of cubic phase of $Co_xMn_{3-x}O_4$ coexists with the dominating tetragonal phase. In our case, however, no such peak splitting is observed in the XRD patterns, suggesting the absence or extremely low fraction of cubic phase could be present. It is worth noting that the detection limit of conventional XRD typically lies around 2–3 wt%, depending on the crystallinity and contrast of the minor phase. So, to detect the existence of a marginal cubic phase of $Co_xMn_{3-x}O_4$, if any, structural analysis has been undertaken through Rietveld refinement considering the tetragonal and cubic phase for the prepared $CoMn_2O_4$. Also, in the spinel structure of $CoMn_2O_4$, a fraction of Co and Mn atoms exchange their positions between tetrahedral (4a) and octahedral (8d) sites, which is also confirmed by the interatomic distance found from EXAFS spectra in literature [41]. To elucidate this partial inversion, the potential incorporation of $Co^{3+}$ into octahedral sites and $Mn^{2+}$ into tetrahedral sites were also considered during the Rietveld refinement process. The refinement of the XRD data was performed using FullProf software and is illustrated in Fig. 1(a), with refined parameter summarized in Table 1. The refinement revealed cation distribution [$(Co_{0.94}Mn_{0.06})_{tetra}(Mn_{1.94}Co_{0.06})_{octa}O_4$] for the tetragonal phase, and [$(Co_{0.04}Mn_{0.94})_{tetra}(Mn_{1.10}Co_{0.90})_{octa}O_4$] for the cubic phase. This suggests the existence of a marginal cubic phase of $Co_xMn_{3-x}O_4$, which may be coexisting with the dominating tetragonal phase. The refined lattice parameters for the tetragonal phase are a = b = 5.721 Å and c = 9.253 Å, with c/a ratio of 1.618. This indicates that the spinel $CoMn_2O_4$ possesses a tetragonal symmetry with I41/*amd* space group (No. 141). The crystal structure of $CoMn_2O_4$ is depicted in the inset of Fig. 1(a). The as grown $CoMn_2O_4$ was subjected to Raman spectroscopy, as depicted in Fig. 1(b). The tetragonal $CoMn_2O_4$ is expected to exhibit ten first order Raman-active modes as T ($B_{1g} + E_g$), L($E_g$), $\nu_1$ ($A_{1g}$), $\nu_2$ ($A_{1g} + B_{2g}$), $\nu_3$ ($B_{1g} + E_g$), $\nu_4$ ($B_{1g} + E_g$) [36, 43]. However, we observe 9 distinct Raman modes which are clearly visible, as shown in Fig. 1(b). The absence of one mode could be attributed to its weak intensity or overlap with adjacent modes. To analyse the spectrum, the Raman pattern was fitted with the Lorentzian curves, and extracted Raman modes align well with the predicted modes. In the low frequency region, peaks P1 and P2 are associated with the stretching vibration of the ($Co^{2+}/Mn^{2+}$) − O bond in the tetrahedral site. In the middle frequency region between 300 and 550 $cm^{-1}$, P3 and P4 correspond to oxygen ligand vibrations in $MnO_6$ and $MnO_4$ sites, while P5 and P6 are related to movements of divalent metal ions in tetrahedral sites. Lastly, in the high frequency region, peaks P7, P8, and P9 are linked to the vibrational modes involving oxygen atom motions within the $MnO_6$ octahedra. These observations confirm the structural integrity of the $CoMn_2O_4$ compound, with Raman modes consistent with the theoretical predictions for a spinel-type structure exhibiting tetragonal distortion. The SEM image and EDX analysis of $CoMn_2O_4$ are shown in Fig. 1(c) and (d), respectively. This is carried out to study the surface morphology and elemental composition. EDX spectrum in Fig. 1(c), illustrates emission peaks of desired elements Mn, Co, and O are present with expected stoichiometry. The experimental atomic and mass percentages (inset of Fig. 1(c)) reveal uniform elemental composition of $CoMn_2O_4$ close to their nominal stoichiometry. The surface micrograph of $CoMn_2O_4$ (Fig. 1(d)), reveals the well-defined, closely packed grains with a predominantly polyhedral morphology typical of spinel structures. However, some intergranular voids or pores are visible that is reflective of some degree of porosity. The grain size distribution appears relatively uniform. Using 'ImageJ' software,

average grain size is estimated as 2.35 ± 0.05 µm from the histogram plots fitted with Gaussian function (inset of Fig. 1(d)).

*3.2. Magnetic Structure Analysis*

To investigate spin-ordering phenomena, temperature dependent magnetic measurements of $CoMn_2O_4$ were conducted under zero-field-cooled (ZFC) and field-cooled (FC) protocols with an applied magnetic field of 200 Oe. The DC magnetization variation with the temperature is illustrated in Fig. 2(a), with the derivative of magnetization, dM(T)/dT, included in the inset. Magnetization measurements exhibit two distinct magnetic transitions, one at temperature $T_1 \sim$ 186 K and another at lower temperature, $T_2 \sim$ 86 K, that are distinctly identified through inflection points in the dM(T)/dT. Below $T_2$, the sublattice and spin structure of the sublattice of $CoMn_2O_4$ are shown in Fig. 2(b) and (c), respectively. The spins of $Co^{2+}$ ions occupying the tetrahedral A-sites align ferromagnetically with the other $Co^{2+}$ cations, and antiferromagnetically with the $Mn^{3+}$ cations, without any frustration. However, the spin of $Mn^{3+}$ at octahedral sites, forming a pyrochlore-like structure (as shown in the left side of Fig. 2(c)), exhibits frustration. Due to competing interactions between $Co^{2+}-Mn^{3+}$ and $Mn^{3+}-Mn^{3+}$, the $CoMn_2O_4$ spinel system adopts an energetically favourable magnetic ground state characterized by a noncollinear triangular spin canting arrangement. This results in a net magnetic moment resulting from the imbalance of spin canting, as is shown in the right side of Fig. 2 (c). This arrangement is described as YK spin structure, in which $Co^{2+}$ spins align along the b-axis, while the two $Mn^{3+}$ spins are canted; one toward the +c-axis, and the other toward the –c-axis, from the –b-axis [44]. Boucher et al. reported the existence of ferrimagnetism with a noncollinear triangular spin canting in $CoMn_2O_4$ at low temperatures, using neutron diffraction analysis [32]. The synthesized $CoMn_2O_4$ also exhibits another high temperature magnetic transition at $T_1 \sim$ 186 K which is assigned to a minor fraction of the cubic phase. Popovic et al. and P. Mahata et al. have reported a similar high temperature transition (~180 K), identifying it as a characteristic feature of bulk $CoMn_2O_4$ due to the formation of a minor impurity (~5%) phase of $Co_xMn_{3-x}O_4$ system [30, 31]. They further observed that this transition diminishes with a reduction in particle size, indicating that the secondary phase associated with $Co_xMn_{3-x}O_4$ is unlikely to develop in smaller $CoMn_2O_4$ particles. Rajput et al. also show that with an increase of x in $Co_xMn_{3-x}O_4$, high temperature magnetic transition ($T_1$) starts to become more prominent while low-temperature magnetic transition ($T_2$) diminishes, ultimately resulting in the exclusive detection of only high temperature magnetic transition ($T_1$) at x = 2 [42]. Hence, the presence of high temperature magnetic transition in our system suggests a marginal impurity phase of $Co_xMn_{3-x}O_4$. To gain more insights into the magnetic behaviour near high temperature transition, the inverse magnetic susceptibility in the paramagnetic region was analysed and fitted using the Curie-Weiss law [45]. The relationship is expressed as

$$\chi = C/(T - \theta_{CW}) \qquad (1)$$

where $\theta_{CW}$ and C are the Curie-Weiss temperature and Curie constant, respectively. From the fitting, as shown in Fig. 3(a), the estimated values of Curie-Weiss temperature ($\theta_{CW}$) and Curie constant are about – 805 K and 0.0379 emu.K/g.Oe, respectively. The significantly large negative value of $\theta_{CW}$ reflects the presence of the frustration of antiferromagnetic interactions of neighbouring atoms or ions, besides the ferrimagnetic character, which sets at $T_1$. Magnetic field-dependent magnetization measurements, are shown in Fig. 3 (b). At 200 K, a linear M–H loop is observed, which indicates the paramagnetic nature of $CoMn_2O_4$. Upon lowering the temperature, a noticeable opening of the hysteresis loop was observed, with a coercivity of 1.71 kOe at 100 K. Coercivity obtained at 70 K, 50 K, and 5 K are 2.16 kOe, 3.31 kOe, and 3.13 kOe, respectively. Such a high coercive field in hysteresis loops indicates the canted spin magnetic ground state. Notably, the persistence of unsaturated hysteresis (M−H) loops in low temperatures, even at a higher magnetic field, strongly indicates the presence of a canted spin configuration, consistent with the YK spin structure present below $T_2 \sim 86$ K. In such a case, there is a linear increase in magnetism that prevents saturation of magnetization even at high fields. Similar to our observation, unsaturation of magnetization is frequently observed in other spinel systems, such as in $CoCr_2O_4$, $Ti_{1-x}Mn_xCo_2O_4$, or $Ni_{1-x}Mg_xCr_2O_4$, with canted or noncollinear spin configurations [46–48]. A magnified version of M−H at 5 K, as shown in the inset of Fig. 3(b). This reveals an asymmetry between the positive coercive (H+) and negative coercive fields (H–) with zero external field cooling from temperature above $T_1$. This suggests the presence of exchange biasing (EB) in $CoMn_2O_4$. This exchange bias effect manifested without applying an external magnetic field during cooling is known as zero-field cooled exchange bias ($H_{ZEB}$), which is quantified as, $H_{ZEB}$ = ($H_+ + H_-$)/2. The $H_{ZEB}$, observed at temperatures 5 K, 50 K, and 70 K are 0.875 kOe, 0.42 kOe, and 0.0874 kOe, respectively. We noted that this exchange biasing behaviour occurs only below $T_2$, which rules out any contribution from the small impurity phase $Co_{3-x}Mn_xO_4$. When the sample is cooled in the absence of a magnetic field up to measurement temperature below $T_2$, the disordered moments of the A and B sublattices remain frozen as in their room temperature configuration. Upon applying a magnetic field, the moments in the tetragonal A sites and moments of the octahedral B site (which are magnetically frustrated) begin to align with the field direction as its strength increases. When the applied field's direction is reversed, the ferromagnetic moments reorient accordingly, while the moments in the B sublattice remain fixed due to their antiferromagnetic interactions. These stationary moments create a microscopic torque on the neighbouring spins in the tetrahedral sublattice that resists the reversal of ferromagnetic spins. This phenomenon can be viewed as the formation of a new interface layer containing both reversible and irreversible spins between the two sublattices, leading to $H_{ZEB}$. The observation of the exchange bias effect in the $CoMn_2O_4$ system offers potential for technological applications [49, 50].

*3.3. Study of Electrical Transport*

To probe the electrical conduction mechanism of $CoMn_2O_4$, temperature dependent variation of DC electrical resistivity ($\rho$) was carried out by the two-probe method in the temperature range 180 to 280 K. This is shown in Fig. 4. The magnitude of resistivity increases from 200 MΩ/cm to 13 GΩ/cm as temperature decreases from 280 K to 180 K, which highlights the electrically insulating nature of $CoMn_2O_4$. In order to understand the transport mechanism, we used the thermal activation model (Arrhenius model), which is represented by the following expression [51],

$$\rho(T) = \rho_0 \, exp \, (E_a / k_B . T) \qquad (2)$$

where $E_a$ is activation energy, $k_B$ is the Boltzmann constant, and $\rho_0$ is constant resistivity. As shown in left inset of Fig. 4, the logarithm of resistivity ln ($\rho$) vs. the inverse of temperature (1/T) exhibits the linear behavior and follows the Arrhenius model in the limit of 212–280 K. Estimated activation energy by fitting the Arrhenius model in the linear region results, $E_a \sim 0.39$ eV. Below the temperature 212 K, ln ($\rho$) vs. the inverse of temperature (1/T) deviates significantly, which indicates the possibility of another transport mechanism. The deviation from the Arrhenius model at low temperatures can be due to a lack of sufficient thermal energy for carriers to occupy higher energy states through free movement. Hence, the conduction mechanism can be understood by hopping of carriers, which is Mott's 3D variable range hopping (VRH) mechanism [52],

$$\rho(T) = \rho_0 \left[exp\,(T_0/T)^{1/4}\right] \qquad (3)$$

where $\rho_0$ is constant resistivity, and $T_0$ is Mott's characteristic temperature, which is dependent on $N(E_F)$ (density of localized states at the Fermi level) and localization length ($\xi$) of charge carrier via relation[53],

$$T_0 = 24[\pi\, k_B\, N(E_F)\, \xi^3]^{-1} \qquad (4)$$

The experimental data is well fitted with Mott's VRH model between the measurable temperature range as inferred in the right inset Fig. 4. This suggests that the conduction mechanism in CoMn$_2$O$_4$ is well followed by the VRH model. The hopping energy $E_h(T)$ at a certain temperature can be determined by the following relation[53],

$$E_h(T) = (k_B T^{3/4}\, T_0^{1/4})/4 \qquad (5)$$

where T is the temperature at which hopping energy needs to be estimated and $T_0$ is Mott's characteristics temperature. The estimated value of $T_0$ through the fitting of ln $(\rho)$ vs. $1/T^{1/4}$) by equation (3) is $(7.47 \pm 0.013) \times 10^9$ K. Furthermore, from equation (5), obtained hopping energy varies from 0.30 eV to 0.42 eV as temperature varies between 180 K and 280 K [54].

*3.4. Study of Dielectric and Magnetodielectric Properties*

The highly insulating nature of CoMn$_2$O$_4$ ensures that observed variations in dielectric permittivity ($\varepsilon$) will be attributed primarily to its intrinsic nature. Fig. 5(a) and (b) represent the dielectric permittivity and loss (tan δ) as a function of temperature at various frequencies. With increasing temperature, a pronounced and non-dispersive change in dielectric permittivity appears approximately at 83 K (Fig. 5(a)). This is reflected by their derivative plotted in the inset Fig. 5(a). This change in dielectric permittivity is observed near the YK spin structural temperature (T$_2$). Moreover, the observation of the extremely small value (0.01) of dielectric loss at low temperature (T < 125 K) suggests the highly insulating nature of CoMn$_2$O$_4$. This eliminates the possibility of contributions from any extrinsic charge carriers, relaxation processes, or experimental artifacts affecting the dielectric permittivity.

On further increase in temperature, the dielectric permittivity exhibited a sharp rise followed by a dispersive behaviour within the range of 150–225 K, accompanied by a peak in dielectric loss as shown in left inset of Fig. 5(b). This is shifted to higher temperatures with an increase in frequency.

In contrast to this behaviour, R. Wang et al. observed the relaxor ferroelectric behaviour in this range of temperature[55]. However, by systematic study of polarization behaviour, it is observed that there is no clear ferroelectric behaviour in the CoMn$_2$O$_4$. In our sample this relaxation phenomenon can be attributed to the thermally activated electron hopping between mixed-valence states, leading to the formation of defect dipoles. Under an applied electric field, the alignment of electron hopping can contribute to macroscopic polarization. This electron hopping induced polarization displays relaxation behaviour, as is evident from the frequency dependent shift of tan δ peaks toward higher temperatures. The relaxation mechanism can be reflected by the Arrhenius model [56],

$$f = f_0 e^{\frac{-E_a}{T_{max}*K_B}} \quad (6)$$

where $f_0$ represents the characteristic frequency, $E_a$ denotes the activation energy associated with dielectric relaxation, and $K_B$ is the Boltzmann constant, $T_{max}$ refers to the temperature at which the dielectric loss shows peak. The right inset of Fig. 5(b) shows ln(f) as a function of $T_{max}$, which follows the Arrhenius model. The evaluated activation energy is $E_a$ ~ 0.26 ± 0.009 eV, which is comparable with activation energy ~ 0.271 eV at a lower temperature (150 K) from the VRH mechanism using the $E_h(T)$ relation (Eq. 5). This relatively low activation energy suggests that polarization in the 150 K–225 K temperature range is primarily governed by localized charge carrier hopping driven by valence state fluctuations in spinel [57, 58]. Additionally, extrinsic factors such as electrode-sample interfaces and grain boundaries effect, significantly enhance the dielectric permittivity of CoMn$_2$O$_4$ at high temperatures (T > 225). To further demonstrate the correlation between magnetic and dielectric properties, temperature dependence dielectric permittivity is measured under 0 and 4 Tesla magnetic fields with frequency of 30 kHz. The dielectric constant under applied magnetic field is shown in Fig. 5(c) with magnified version in its inset. In the low temperature range (lower inset of Fig. 5(c)), the dielectric permittivity is suppressed with the application of a magnetic field. This suppression is maximum below the temperature where magnetic transition T$_2$ and dielectric permittivity coexist, which indicates a magnetodielectric effect. However, the dielectric permittivity enhanced significantly under a magnetic field at a higher temperature as shown in the upper inset of Fig. 5(c). This opposite effect could be due to activation of electron hopping, which will be influenced by the magnetic field through the Lorentz force or the combination of the magnetoresistance and Maxwell-Wagner effect. To further validate the strength of magnetodielectric (MD) coupling, isothermal MD (%) (defined as [ ε (H) – ε (H = 0) / ε (H = 0)] × 100) was measured at 30 kHz as function of magnetic field at different temperatures. The change in dielectric permittivity as a function magnetic field at 50 K and 70 K is depicted in the inset of Fig 5(d). The change in the MD (%) is of the order of $10^{-2}$ (0.04% at 70 K and 0.023% at 50 K), which is comparable to that in M-type hexaferrites and other spinels [59–61]. To understand the MD coupling in CoMn$_2$O$_4$, a theoretical framework based on the Ginzburg-Landau theory is employed, which provides a phenomenological approach to describe second-order phase transition. According to this approach, thermodynamic potential ($\Phi$) can be expressed in the terms of magnetic, electric, and the magnetoelectric coupling terms $\gamma P^2 M^2$ as[61],

$$\Phi = \Phi_0 + \alpha P^2 + \frac{\beta}{2} P^4 - PE + \alpha' M^2 + \frac{\beta'}{2} M^4 - MH + \gamma P^2 M^2 \tag{7}$$

Where α, α', β, β', and γ are temperature dependent magnetodielectric coupling coefficients, while P and M are the polarization and magnetization, respectively. The external magnetic and electric fields are denoted by E and H. From the above expression, the dielectric permittivity can be obtained by taking the second derivative of $\Phi$ in terms of P as

$$\varepsilon = \chi_e + 1 = \frac{1}{\frac{\partial^2 \Phi}{\partial P^2}} + 1 = \frac{1}{\alpha + \gamma M^2} + 1 \tag{8}$$

$$\varepsilon \approx 1 + \frac{1}{\alpha}\left(1 - \frac{\gamma}{\alpha} M^2\right) \tag{9}$$

From this general formalism, the difference of the relative dielectric permittivity is proportional to the square of the magnetization. Magnetodielectric coupling (Δε %) is plotted as a function of $M^2$ at 70 K in Fig. 5(d). The linear agreement between Δε (%) and $M^2$ provides compelling evidence that the magnetodielectric coupling originates from the term $\gamma P^2 M^2$ of Ginzburg-Landau framework in the spinel CoMn$_2$O$_4$. This observation aligns with similar findings in other spinel and oxide systems [61, 62].

*3.5. Study of Ferroelectric Aspects*

In order to clarify whether there is any intrinsic ferroelectric order in the compound, we carried out a detailed study of temperature dependent pyroelectric current Ip(T). The intrinsic ferroelectric order is typically characterized by a spontaneous electric polarization when the structural transition occurs from a non-polar phase to a polar phase. The charges accumulated at the material surface due to spontaneous polarization will be released upon reaching the transition temperature from low temperatures, which results in the detection of the pyroelectric current with appearance of a peak, which is independent of the rate of heating and poling temperature [63, 64]. For $CoMn_2O_4$, the pyroelectric current as a function of temperature is shown in Fig. 6. For the pyroelectric current measurements, the sample was initially poled from the high temperature (~ 230 K) to the low temperature while subjected to an external electric field of ±3.1 kV/cm. To remove spurious charge accumulation on the surface during poling, the sample was shorted for about 40 to 45 minutes. Then, pyroelectric current was recorded during warming with a controlled heating rate of 5 K/min. The observed pyroelectric current shows two anomalies, a small anomaly at ~83 K and a large anomaly at ~195 K temperature, as depicted in Fig. 6(a). The inset of Fig. 6(a) shows clear anomaly near ~81 K with poling temperature of 185 K. The presence of a pyroelectric peak can arise from both intrinsic and extrinsic effects, so to distinguish extrinsic contribution, we performed different experimental protocols while measuring the pyroelectric current [65]. Firstly, as shown in Fig. 6(b), pyroelectric current was measured with the different warming rates (2, 3, 4, and 5 K/min), keeping the poling temperature fixed at 230 K. Generally, the stable pyroelectric peak independent of heating rate indicates the intrinsic ferroelectric transition in materials such as in $Bi_2Fe_4O_9$ and $MnCr_2S_4$ [66, 67]. However, our results indicate that pyroelectric peaks are affected by the heating rate, which indicates that there is no first-order ferroelectric transition in $CoMn_2O_4$. To further confirm whether the emergence of the pyroelectric peak is related to ferroelectric order or not, electric field-dependent polarization (P(E)) measurements were performed at different temperatures (Fig. 6(c). The linear nature of the P(E) response is characteristic of dielectric behaviour, indicating the absence of any spontaneous polarization.

To verify the thermal activation of the pyroelectric peak, pyroelectric current was measured with a fixed warming rate (5 K/min) after cooling the sample from different poling temperatures, $T_{pole}$, (Fig. 6(d)). Both the pyroelectric peaks shifted towards the lower temperature with lower poling temperature and their intensity significantly weakened. This observation indicates that poling temperature significantly influences the stability of the pyroelectric current peak position, and the observed peak can be attributed to thermally stimulated depolarization current (TSDC) [68]. TSDC arises from mobile charge carriers associated with point defects, which align under an applied electric field and become trapped as the temperature decreases, forming stable electric dipoles.

Upon reheating, these trapped carriers are thermally released, generating peaks in pyroelectric current measurements. Recently, TSDC peaks have been observed in a large number of systems [64, 68]. Therefore, pyroelectric behaviour observed in $CoMn_2O_4$ is attributed to the thermally stimulated depolarization effect rather than the intrinsic ferroelectricity. The occurrence of two distinct pyroelectric peaks has previously been reported in $CoMn_2O_4$, within the temperature range of approximately 70–130 K [55]. Similar behaviour is observed in our current work, except that the higher temperature pyroelectric peak, as observed by R. Wang et al., appears to be merged with a broad anomaly around 195 K in our sample. R. Wang et al. attributed these pyroelectric features to the thermally activated response of defect dipoles and electric domains. Both effects give rise to TSDC peaks which is consistent with our current finding. However, non-dispersive change in dielectric permittivity appears at temperature ~83 K, which is close to the YK spin structure. This confirms the presence of magnetodielectric coupling in $CoMn_2O_4$.

Recently, spin-driven ferroelectricity by the YK structure in $MnCr_2S_4$ through the magnetostriction effect has been reported [67]. $CoMn_2O_4$ spinel has the YK structure, yet no evidence for intrinsic ferroelectricity could be found. However, the absence of ferroelectricity in $CoMn_2O_4$, does not rule out magnetodielectric coupling. A similar scenario has been observed in other systems, $Mn_3O_4$, $Ni_2Te_3O_8$, and $Ni_2ScSbO_6$, which exhibit the MD effect combined with spin-phonon coupling [62, 69, 70] without any ferroelectric phase. The MD coupling in $CoMn_2O_4$ could be assigned to spin-phonon coupling [36].

## 4. Conclusion

In summary, we have investigated the magnetic, dielectric, and ferroelectric characteristics in polycrystalline spinel $CoMn_2O_4$. We find no evidence of intrinsic ferroelectricity in this compound. Two magnetic transitions are observed. The co-dependence of dielectric property as a function of magnetic field is observed across the ferrimagnetic transition temperature ~86 K. Pronounced magnetodielectric coupling is seen below the ferrimagnetic transition. Another high temperature magnetic transition around 186 K, is assigned to formation of minor secondary cubic phase. Magnetodielectric coupling is not seen across this transition. Raman studies establish strong spin-phonon coupling in the extended temperature range. A significant exchange bias effect in observed in $CoMn_2O_4$. The variation of dielectric permittivity follows square of magnetization, which supports the applicability of Ginzburg-Landau theory to explain magnetodielectric coupling. The microscopic origin for this phenomenon of coexisting magnetodielectric coupling via spin-phonon coupling relates to Yafet-Kittel type spin structure.

**Acknowledgment**

P. Kumar sincerely thanks CSIR, India, for awarding the Junior Research Fellowship (JRF), which provided financial support for this work. P. Das acknowledges the financial support received through UGC-NET JRF. We sincerely appreciate and thankful to the Department of Science and Technology's FIST program, Government of India, for facilitating access to the low-temperature, high-magnetic-field measurement facility at JNU. Additionally, we acknowledge the CIF of SPS and SCNS, JNU, for providing the structural characterization facility. SP thanks support from DST Nano mission project DST/NM/TUE/QM-10/2019 (G)/6. We are also thankful to the Advanced Instrumentation Research Facility at JNU for granting access to the PPMS facility.

**Conflict of Interest**

All authors declare that they have no financial/commercial conflicts of interest.

**Data Availability Statement**

The data that support the findings of this study are available from the corresponding author upon reasonable request.

**Figure Captions**

Figure 1: (a) Refined XRD pattern of tetragonal $CoMn_2O_4$ spinel is depicted. The schematic crystallographic unit cell is included in the inset. The blue, cyan, and black circles show Co, Mn, and O atoms, respectively. (b) The room temperature Raman spectra of $CoMn_2O_4$. (c) EDX spectra of $CoMn_2O_4$. Its inset shows the experimental atomic and mass percentage of elements in tabular form. (d) SEM image of $CoMn_2O_4$. Inset shows the distribution of particle size.

Figure 2: (a) DC magnetization curves measured with 200 Oe are shown. Inset displays the derivative of M(T). to clarify the magnetic transition temperature. (b) Local coordination in $CoMn_2O_4$ with $Co^{2+}$ occupying the tetrahedral A site and $Mn^{3+}$ occupying the octahedral B site, corresponding to the unit cell as shown in the inset of Fig(1a). (c) The schematic view of noncollinear spin configuration (right) having competing interactions that induce the Yafet-Kittel ferrimagnetic phase of $CoMn_2O_4$ (left). Lattice unit vectors, a, b and c are shown. The blue and cyan arrows represent the spins of Co and Mn cations, respectively.

Figure 3: (a) Inverse of magnetic susceptibility with Curie-Weiss analysis. (b) Isothermal magnetization as a function of external magnetic field is plotted. Inset of (b) shows Magnified M-H loop at 5 K that establishes exchange bias in the system.

Figure 4: Temperature dependence of resistivity of $CoMn_2O_4$ spinel. The inset shows the Arrhenius model (left) and Mott's VRH (right) fitted curves.

Figure 5: (a) and (b) show temperature-dependent $\varepsilon'$ and loss tangent at different frequencies. The inset of (a) shows the derivative of $\varepsilon'$. The left inset of (b) shows magnified view of dielectric loss in the temperature range of 125 to 250 K. The right inset of (b) shows ln(f) vs. 1000/T fitted by the Arrhenius model. (c) shows $\varepsilon'(T)$ under 0 and 4 T magnetic fields, and Insets of (c) represent the magnifications of curves in the vicinity of low and high temperatures. (d) Plot of $M^2$ with $|\Delta\varepsilon'| = [(\varepsilon'(H)/\varepsilon'(0)) - 1]$ (%) at frequency of 30 kHz. The straight line depicts the linear fit at temperature 70 K, and its inset shows the change in magnetodielectric ($\Delta\varepsilon'$) at 50 and 70 K temperatures.

Figure 6: (a) Pyroelectric current ($I_p$) for different reverse poling voltage (±3.1kV/cm) with poling temperature 230 K. Inset shows the magnified version of pyroelectric current ($I_p$) with poling temperature 185 K and poling voltage +3.1kV/cm. (b) Pyroelectric current with different warming rates by keeping constant poling temperature 230 K. (c) P-E loop measurements up to $E_{max}$ = 8 kV/cm at 200 K, 150 K, and 80 K, respectively. (d) $I_p$ with different temperatures $T_{pole}$ with constant heating ramp 5K/mint. Insets of (a), (b), and (c) show magnified version of pyroelectric current at low temperatures.

Table 1: Refined structural parameters of $CoMn_2O_4$ with the Wyckoff positions from the Rietveld refinement.

**Tetragonal Phase**
**Lattice parameters:** a = b = 5.721 Å, c = 9.253 Å

| Atom | Site | X | Y | Z | Site occupancy |
|---|---|---|---|---|---|
| Co (A-site) | 4b | 0 | 0.25 | 0.375 | 0.94 |
| Mn (A-site) | 4b | 0 | 0.25 | 0.375 | 0.06 |
| Co (B-site) | 8c | 0 | 0 | 0 | 0.03 |
| Mn (B-site) | 8c | 0 | 0 | 0 | 0.97 |
| O | 16h | 0 | 0.523 | 0.245 | 1.00 |

**Cubic Phase**
**Lattice parameters:** a = b = c = 8.278 Å

| Atom | Site | X | Y | Z | Site occupancy |
|---|---|---|---|---|---|
| Co (A-site) | 4b | 0.125 | 0.125 | 0.125 | 0.04 |
| Mn (A-site) | 4b | 0.125 | 0.125 | 0.125 | 0.94 |
| Co (B-site) | 8c | 0.5 | 0.5 | 0.5 | 0.45 |
| Mn (B-site) | 8c | 0.5 | 0.5 | 0.5 | 0.55 |
| O | 16h | 0.262 | 0.262 | 0.262 | 1.00 |

**Figure 1.**

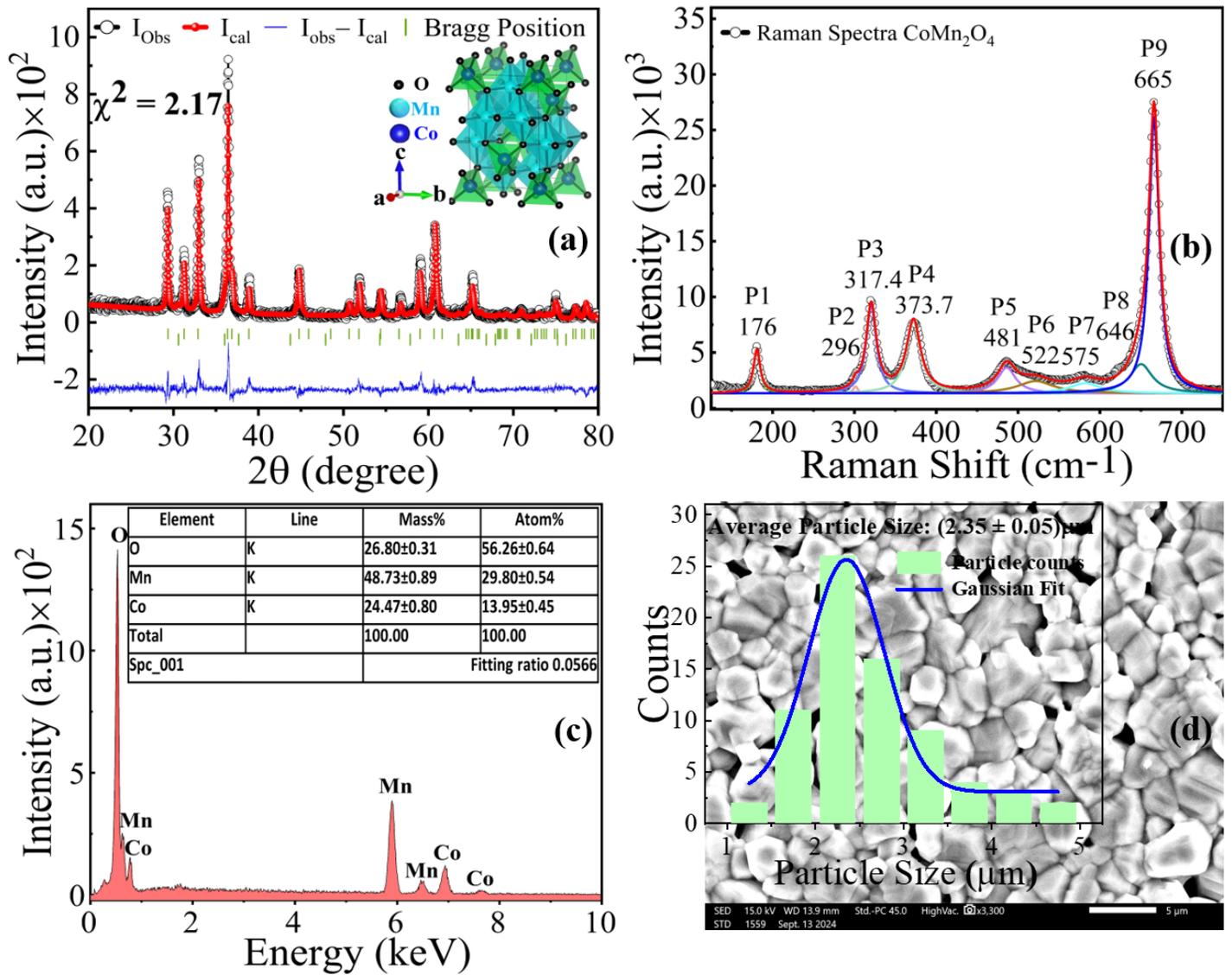

**Figure 2.**

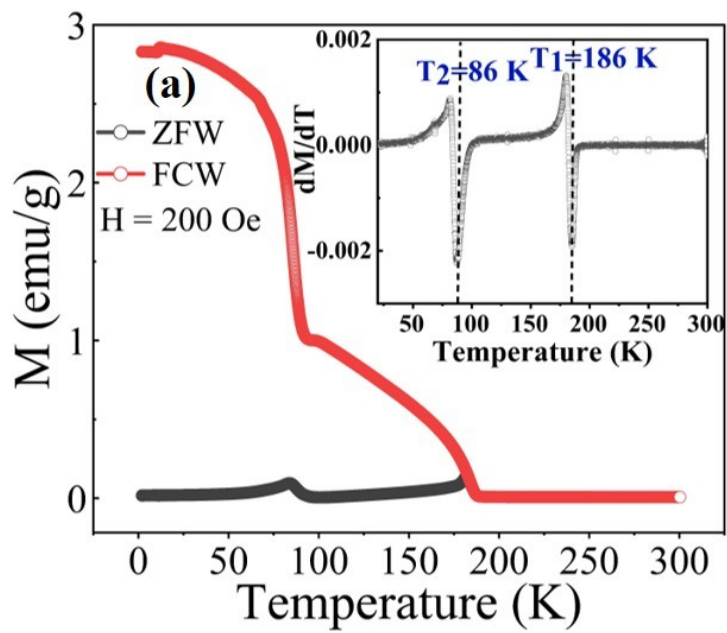

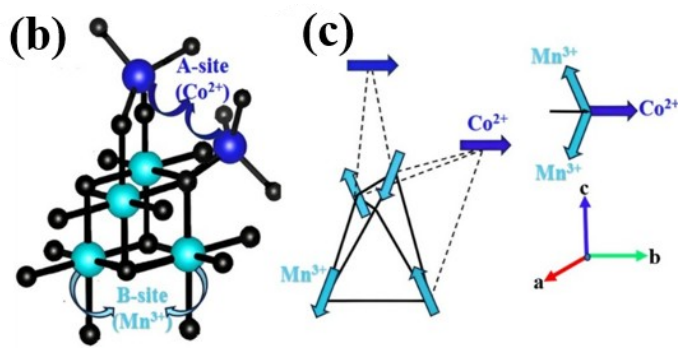

**Figure 3.**

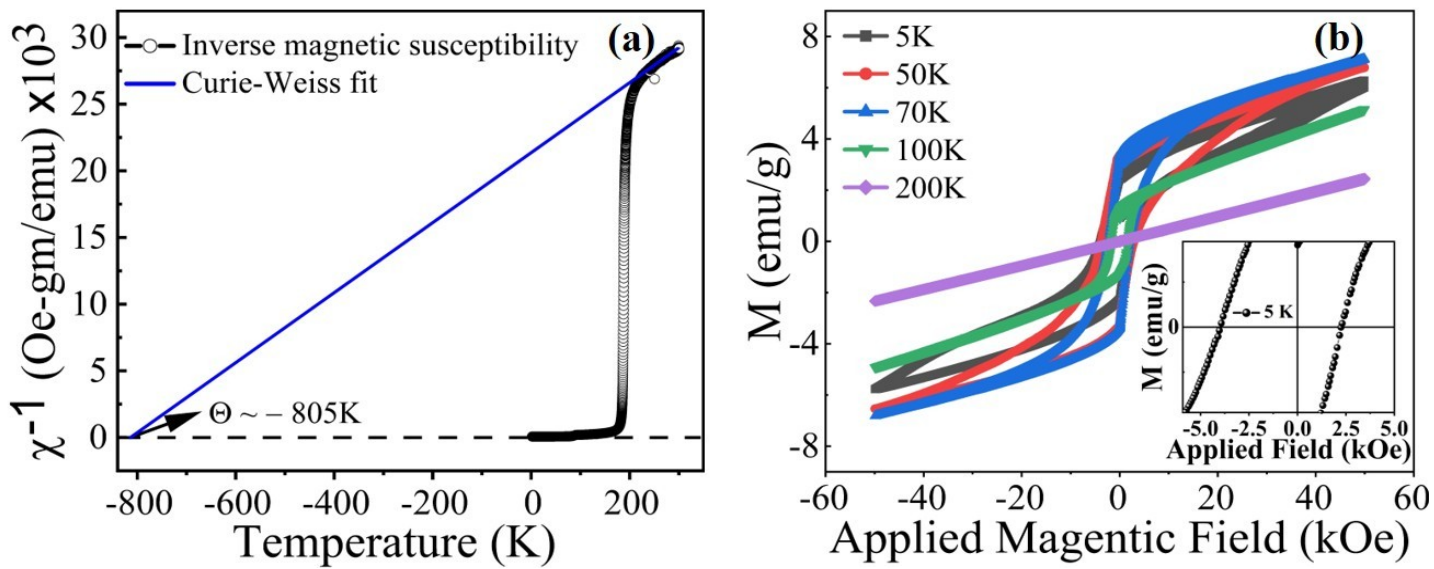

**Figure 4.**

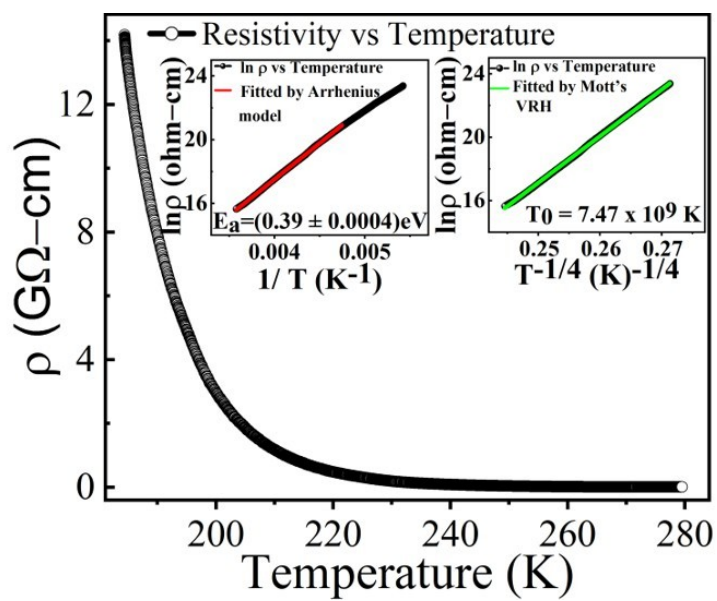

**Figure 5.**

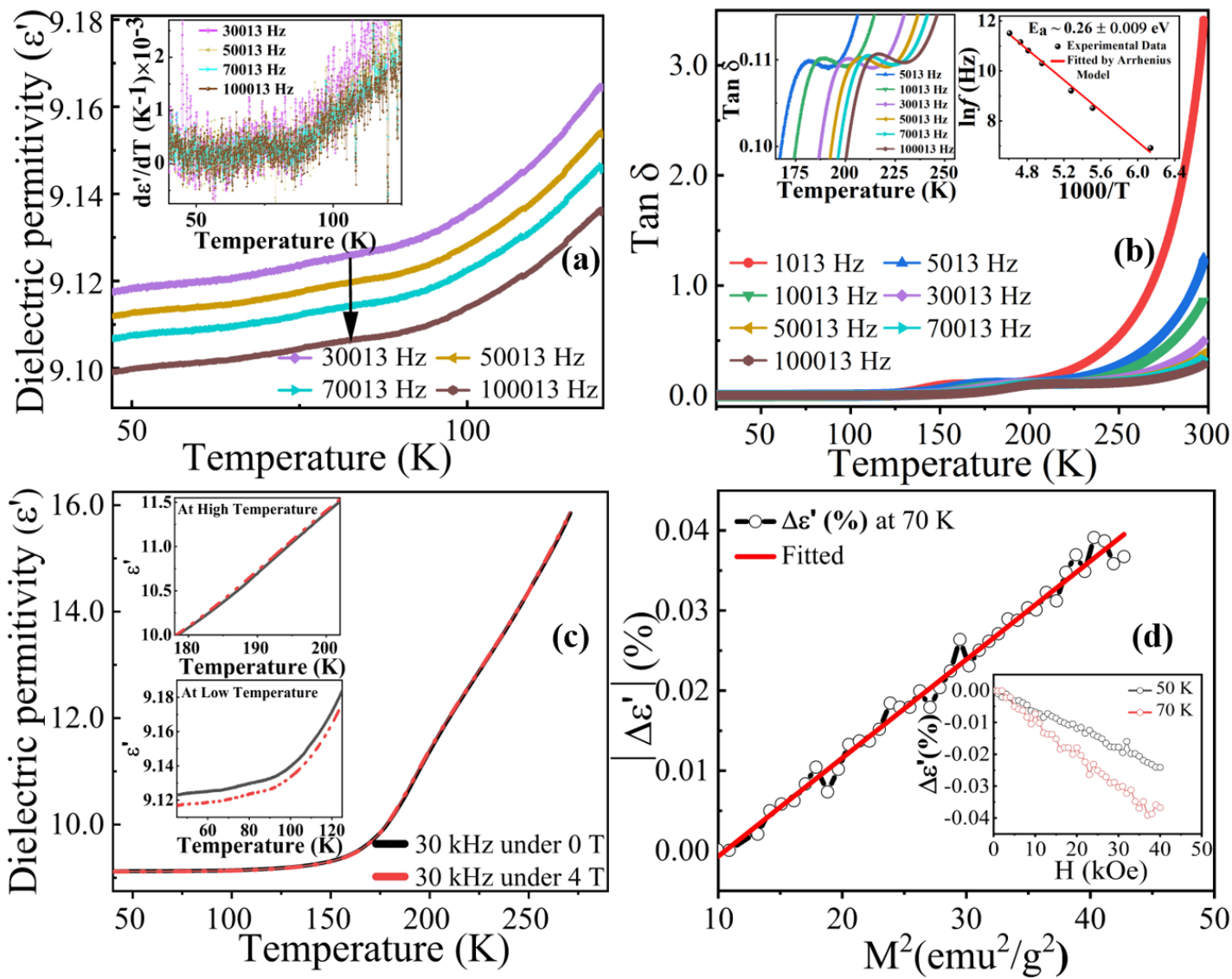

**Figure 6.**

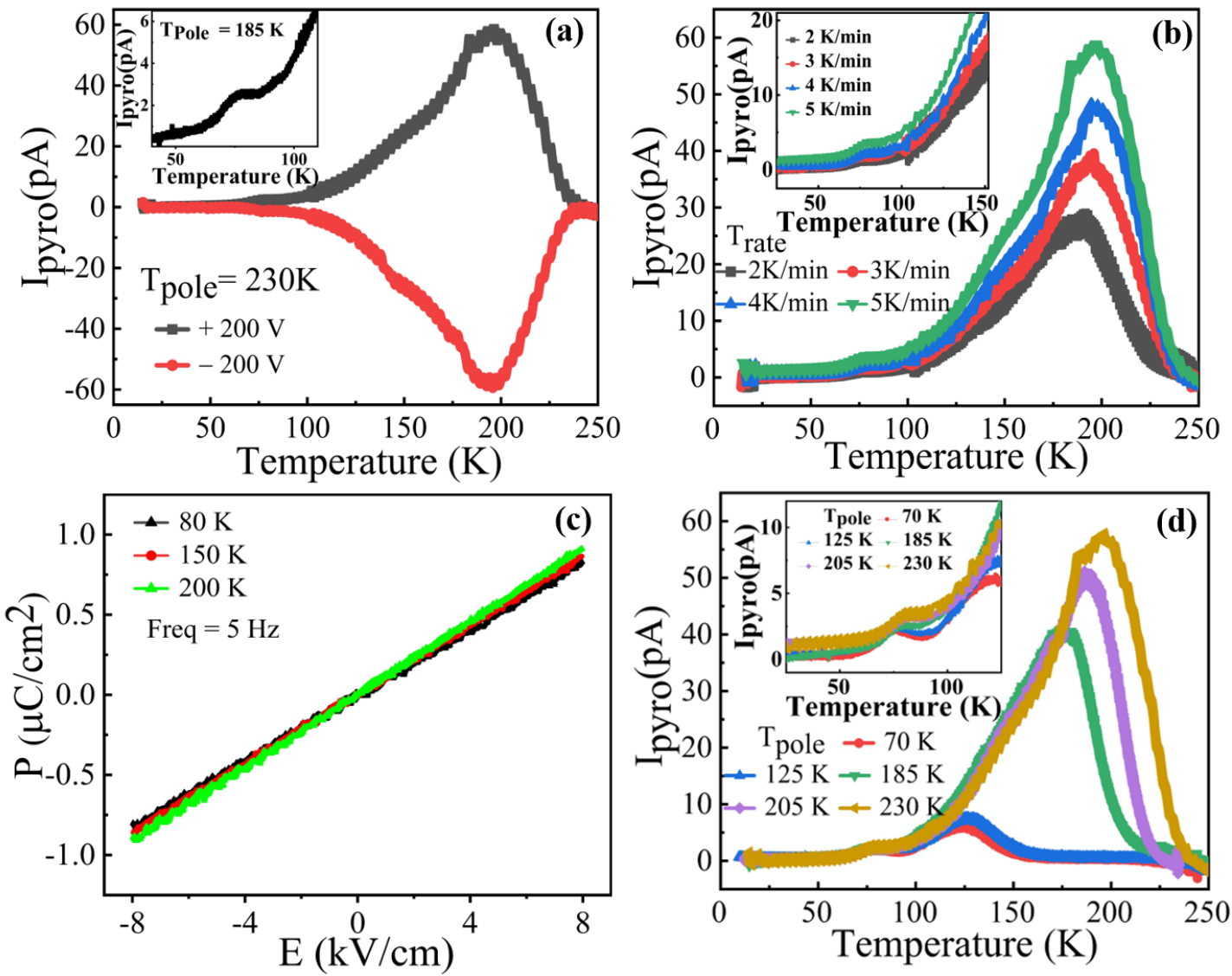